\documentclass[journal, letterpaper]{IEEEtran}
\usepackage{amsmath}
\usepackage{graphicx,subfigure,comment}
\usepackage[table]{xcolor}
\usepackage{url} %
\usepackage{dsfont}
\usepackage{cite}
\usepackage{tabularx}
\usepackage{dblfloatfix}

\newcommand{\Fig}[1]{Fig.~\ref{fig:#1}}

\newcommand{\Sec}[1]{Sec.~\ref{sec:#1}}
\newcommand{\Tab}[1]{Tab.~\ref{tab:#1}}

\begin{document}

\title{Towards Node Liability in Federated Learning:\\Computational Cost and Network Overhead}
\author{Francesco Malandrino, Carla Fabiana Chiasserini
\thanks{This work was supported through the EU 5Growth project (Grant No. 856709).}
} %
\maketitle

\begin{abstract}
Many machine learning (ML) techniques suffer from the drawback that their output (e.g., a classification decision) 
is not clearly and intuitively connected to their input (e.g., an image). To cope with this issue, 
several {\em explainable ML} techniques have been proposed to, e.g., identify which pixels of an input image 
had the strongest influence on its classification. However, in distributed scenarios, it is often more important 
to connect decisions with the information used for the model training and the nodes supplying such 
information. To this end, in this paper we focus on federated learning and present a new methodology, named 
{\em node liability in federated learning} (NL-FL), 
which permits to identify the source of the training information that most contributed to a given decision.
After discussing NL-FL's cost in terms of extra computation, storage, and network latency, we demonstrate its 
usefulness in an edge-based scenario. We find that NL-FL is able to swiftly identify misbehaving nodes and 
to exclude them from the training process, thereby improving learning accuracy.
\end{abstract}

\section{Introduction}
\label{sec:intro}

Originally introduced in~\cite{konen2015federatedOptimization}, federated learning (FL) is one of the most relevant 
approaches to the task of distributed machine learning (ML). Its main advantage is that the participating 
{\em learning nodes} can cooperatively train a single ML model (usually a deep neural network, DNN) 
through an iterative procedure, 
without sharing their local data, which may be private and/or sensitive.

\begin{figure}
\centering
\includegraphics[width=.7\columnwidth]{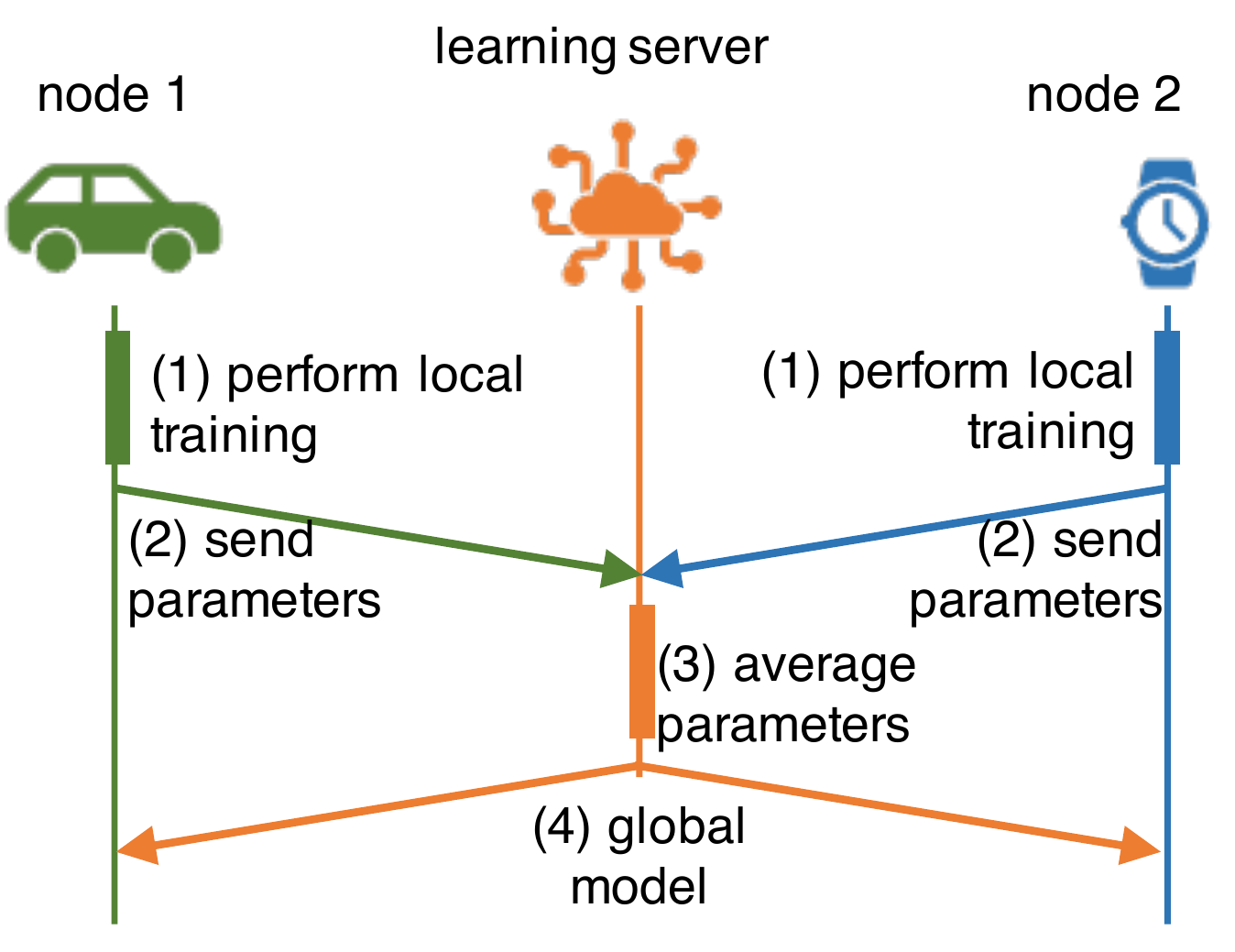}
\caption{
    The main steps performed at each training epoch in FL.
    \label{fig:steps}
} %
\vspace{-.5cm}
\end{figure}

As summarized in \Fig{steps}, each iteration of the FL paradigm, also called epoch, 
consists of the following main steps. First, 
 each learning node trains its local model using its own data (step~1 in \Fig{steps}), and sends the local model 
 parameters to a {\em learning server} (step~2), 
with the latter being often located at an edge host in a mobile network scenario~\cite{wang2019adaptive,client-selection}.
The learning server  averages the received parameters (step~3), and sends 
 the averaged model back to the learning nodes (step~4), which use it in the subsequent epoch.

FL is a very popular option in scenarios where nodes
belong to different individuals and/or administrative entities, as it is the case for user 
personal devices~\cite{konen2015federatedOptimization} or fog nodes~\cite{tran2019federated,infocom20-fog}.
In such cases, nodes often exchange the data messages depicted in \Fig{steps} through wireless 
networks belonging to a third party. 
However, limited and/or unreliable connectivity may negatively impact learning performance~\cite{tran2019federated,infocom20-fog,jeong2018communication}.

Typically, a decision made through the trained model during the inference phase, 
e.g., the classification of an image, 
contains no indication on how it was obtained, e.g., (i) based on which pixels of the image itself, and/or (ii) 
which images from the training set have most contributed to such decision.
Analyzing the latter aspect allows establishing a link between decisions (e.g., a self-driving car failing to recognize a highway barrier, or 
Microsoft's Tay chatbot bringing up conspiracy theories) and the {\em source} of the training data determining 
such decisions (e.g., which crowd-sourced images or tweets).
The same issue is present
in all cooperative learning scenarios, where it is of paramount importance to identify the learning nodes
that, due to malicious behavior 
or simple malfunctioning, inject incorrect information in the learning process.

In this paper, we tackle this issue by introducing the methodology of {\em node liability in federated learning} (NL-FL).
NL-FL aims at identifying
the learning nodes that provided, during the {\em training phase}, the information that had the 
strongest influence over a given decision, e.g., a wrong manuever by a self-driving car.
To this end, NL-FL leverages results from layer-wise 
relevance propagation~\cite{bach2015pixel} and techniques from
model weighting and client selection in FL~\cite{client-selection}. 
Once the misbehaving learning nodes are identified, they can be removed from the learning process; 
further action (e.g., a criminal investigation) can be taken if warranted.

In order to be suitable to fog networking scenarios, with fleeting connectivity and energy constrained 
nodes, NL-FL neither  leads to additional network overhead compared to traditional FL, 
nor requires additional operations at the learning nodes. Indeed, all tasks are performed by the learning server, which is best equipped to perform them.

The remainder of this paper is organized as follows.
First, \Sec{relwork} reviews state-of-the-art works
in the areas of FL client selection, explainable ML, and layer-wise relevance propagation.
Then  \Sec{nlfl} introduces NL-FL, 
describing its underlying principles and implementation details. To showcase the effectiveness of NL-FL, 
we assess its ability to identify the misbehaving nodes in a realistic FL scenario, 
as detailed in \Sec{results}. 
Finally, in \Sec{discussion} we discuss the outcome of our experiments and sketch open challenges calling 
for further research, before concluding the paper in \Sec{conclusion}.

\begin{figure*}
\centering
\includegraphics[width=.65\textwidth]{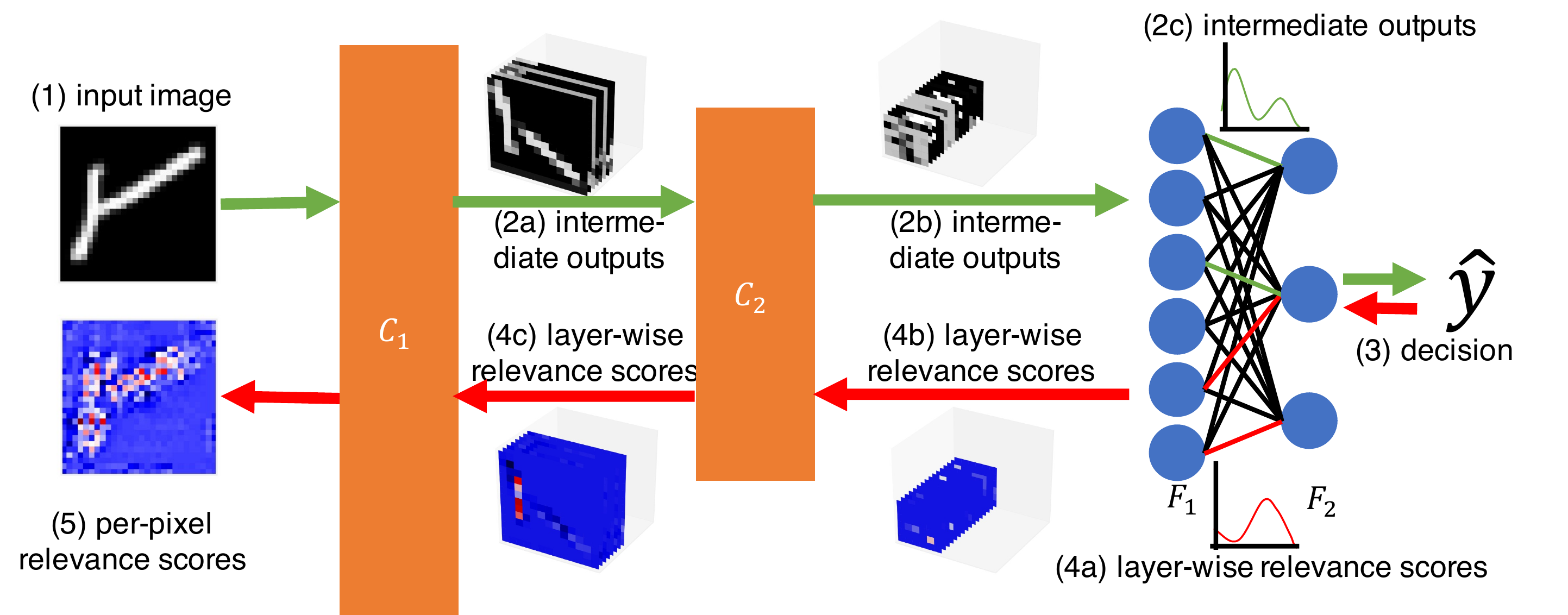}
\caption{
Relevance computation during the inference phase: Schematic view of how a simple DNN processes input images to reach a decision (top, green arrows), 
and how relevance is propagated across layers (bottom, red arrows).
Inputs and outputs of fully-connected layers (right) are one-dimensional (``flat'') tensors, hence, 
they are represented by their empirical pdf.
\label{fig:fresco}
} %
\end{figure*}

\section{Current issues in FL management}
\label{sec:relwork}

Managing a large-scale, distributed learning process like FL is challenging for several reasons, 
including (i) the need to select the learning nodes to rely upon and properly weight the information 
they provide, (ii) the need to {\em explain} the decisions reached, 
and (iii) the need to establish the {\em relevance} of different parts of the input  
on the overall decisions. Below, we discuss how each of these  aspects has been dealt 
with in the literature.

\subsection{Client selection and model weighting in FL}
\label{sec:sub-client}

As DNNs require large datasets to be trained properly, 
the ability of individual learning nodes to contribute to the global learning 
task strongly depends upon 
the quantity of data at their disposal. To account for this, the learning server 
can make the weights assigned to local models 
 in the averaging phase 
 (step~3 of \Fig{steps}) proportional to the size of local datasets, as done 
 in~\cite{konen2015federatedOptimization,tran2019federated}.

Along with the quantity of data available to learning nodes, its quality,
e.g., whether or not it is i.i.d., 
has a very important impact on the 
training performance.

Three main approaches exist to tackle low-quality local data: (i) data augmentation, e.g., combining local and 
remote data samples to obtain new samples~\cite{jeong2018communication,shin2020xor}; (ii) selecting i.i.d. subsets 
from local datasets~\cite{infocom20-noniid}; (iii) weighting local models according to a measure of data quality, 
e.g., entropy.

Besides their local datasets, learning nodes can have different computational capabilities and/or different ways 
to connect to the edge server.

One approach, followed in~\cite{tran2019federated,infocom20-fog}, is to assign more resources (e.g., radio resource blocks) 
to the nodes that need them the most (e.g., experience connectivity issues), so as to avoid performance bottlenecks. 
Another option is simply to drop overly-slow nodes (``stragglers'') from the learning process, 
thus making individual iterations faster~\cite{neglia,tran2019federated,client-selection}. 

Finally, not all clients 
always behave correctly. %
A node may create multiple identities to 
influence the learning process (Sybil attack), or it may 
send incorrect updates to the learning server. As discussed in~\cite{fung2020limitations},
such nodes can be identified by leveraging the notion of {\em distance} between local and global models: 
consistently high distances suggest malfunctioning, while too-close local models coming from different nodes can reveal a Sybil attach.
An alternative approach is presented in~\cite{kang2020reliable}, leveraging blockchain technology to build a reputation system for learning nodes.

\subsection{Explainable ML}
\label{sec:sub-explain}

The high-level goal  of Explainable ML (XML) is to make the decisions of ML systems understandable by humans. This is often achieved by spelling out which elements of the input have had the most significant impact on the decision itself, e.g., ``the application for a credit card was denied because the income was lower than a threshold''.

ML techniques based on decisions trees~\cite{quinlan1990decision} naturally lend themselves to XML, and 
have long been the most popular option whenever the right to explanation has to be guaranteed. 
Similar to flow charts, the root and intermediate nodes of decision trees represent conditions against which 
input data can be checked  (e.g., whether the amount of a transaction is above a threshold); 
the leaves correspond to the ML decision, e.g., whether the transaction is deemed a fraud. 
Decision tree learning algorithms seek to optimize the order and content of the rules, so as to maximize the 
learning quality.

On the negative side, not all types of data are suitable for decision trees,
and many ML applications leverage DNNs instead.

Due to their relevance and popularity, several techniques have been developed in order to explain DNN decisions; 
among such techniques, the most promising one is layer-wise relevance propagation, discussed next.

\subsection{Layer-wise relevance propagation}
\label{sec:lwr}

The high-level goal of layer-wise relevance, introduced in~\cite{bach2015pixel}, is to explain DNN decisions. Considering as an example the image classification task exemplified in \Fig{fresco}, we want to associate a {\em relevance score} (step~5) to each pixel of the original image (step~1), expressing how important each pixel was in reaching the decision. In the figure, we can observe that pixels corresponding to {\em transitions} between dark and light parts of the image tend to have high relevance, while pixels on the background have low relevance: recalling that shapes are defined by transitions between light and dark areas of an image, it makes intuitive sense that those areas matter the most for shape classification.

Relevance values are computed according to the rules laid out in~\cite{Montavon2019}: 
we start from the neuron in the output layer associated with the highest score (i.e., corresponding to the final decision), 
and traverse the DNN back towards the input. At each step, layer-wise relevance values 
express how strongly each element of the previous layer has influenced each element of the current one. 
Importantly, layer-wise relevance values (steps~4a--4c in \Fig{fresco}) have the same shape as the corresponding 
intermediate outputs (steps~2a--2c in \Fig{fresco}). 
This holds also for the  one-dimensional tensors that are inputs and outputs of the fully-connected layers on the right, 
which are represented by their probability density functions (pdf).
As a result, the final relevance scores (step~5 in \Fig{fresco}) can be mapped to individual pixels of the input image.

Crucially, relevance values are computed during the {\em inference} phase, and do not require any change to the 
training phase.

\subsection{Motivation and novelty}

NL-FL is a variant of the FL paradigm, and it works exactly like ordinary FL when no issues are present, i.e., when all nodes behave correctly. When misbehaving nodes are present, NL-FL seeks to identify and exclude them in a similar way to the client selection techniques described in \Sec{sub-client}; however, it is able to only focus on a subset of {\em specific decisions}, e.g., racist tweets by Microsoft Tay. This allows NL-FL to better deal with sophisticate attackers, aiming to only sway some decisions as opposed to sabotaging the whole learning process. At the same time, nodes providing different updates due to their local datasets are not suspected.

The purpose of NL-FL is similar to the XML techniques discussed in \Sec{sub-explain}, in that both seek to explain ML decisions. However, XML focuses on which parts of the input (e.g., which pixels of an image) influenced a given decision (``this guard-rail was identified as a road line because of these white traces on it''). Conversely, NL-FL aims at identifying the {\em training-time} data such decisions are based upon, and their source (``user CrackMonkey74 tagged these photos of guard-rails as road lanes'').

NL-FL reaches its goal by combining the relevance techniques reviewed in \Sec{lwr} with information on the difference between local and global models, as set forth next.

\begin{figure}
\centering
\includegraphics[width=.9\columnwidth]{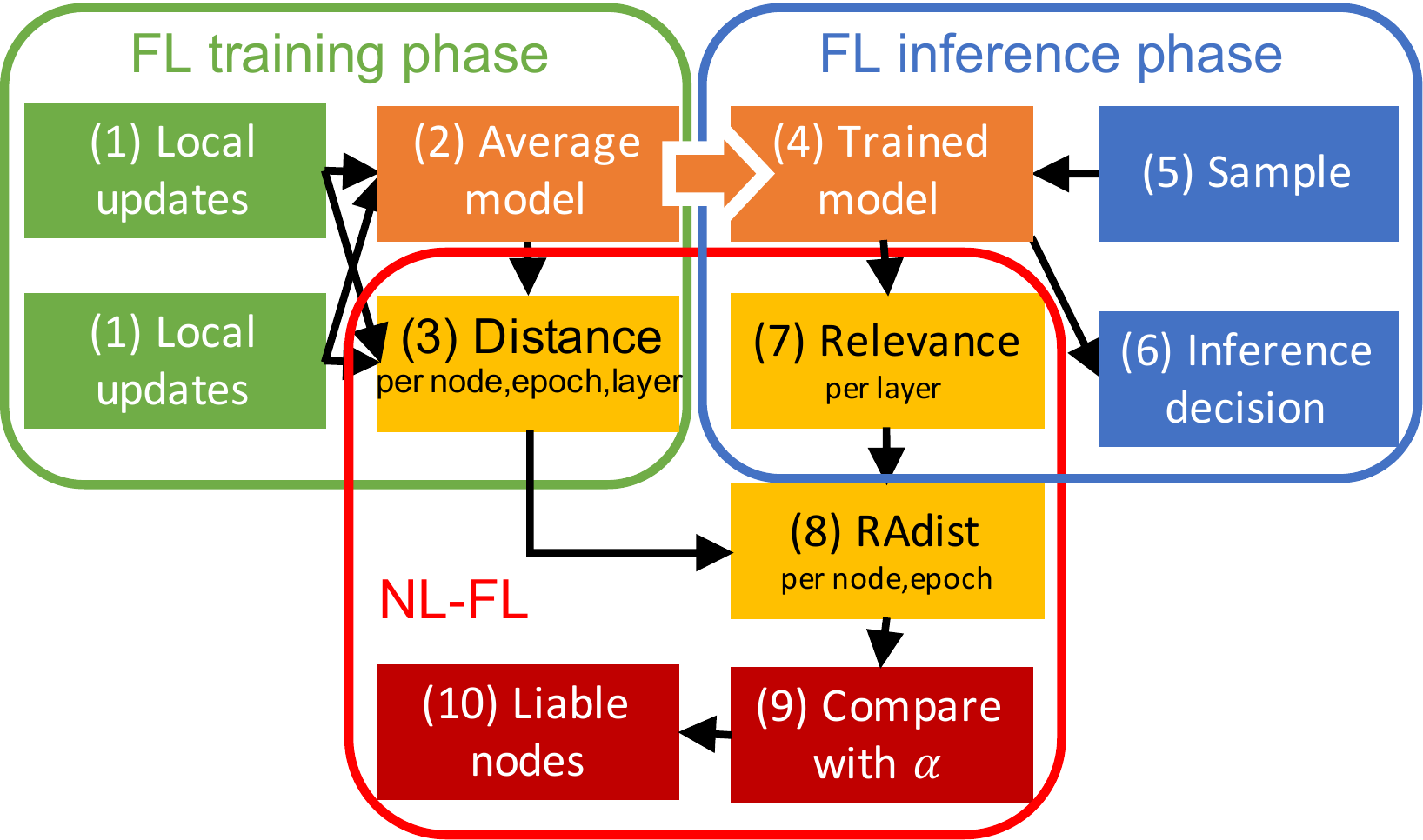}
\caption{
Operations during the training (green) and inference (blue) phases of the FL paradigm, 
and of the NL-FL (red). 
\label{fig:nlfl}
} %
\end{figure}

\section{Node liability in federated learning}
\label{sec:nlfl}

We now introduce the proposed NL-FL technique that seeks to establish how much influence individual 
learning nodes (and their data sets) have on a given decision during the inference phase.

\Fig{nlfl} summarizes the operations carried out by the FL paradigm and the 
additional ones envisioned by our proposed NL-FL. Specifically, the green box in \Fig{nlfl} 
corresponds to the training phase of FL, where learning nodes send local updates to the learning server, 
and the latter computes the average (global) model (see also \Fig{steps}). 
The blue box  denotes the inference phase, where the trained (global) model 
is used to make decisions (e.g., classification) over individual samples. 
The red box includes the operations performed as a part of our NL-FL methodology. 
Numbers in the box identify the chronological order of each step.

During the training phase, the learning server
leverages the information sent by learning nodes as a part of the FL methodology to keep
a log of how similar the updates coming 
from learning nodes at each epoch are to the average model. In particular,
the learning server computes the cosine distance~\cite{fung2020limitations} between the local 
updates and the average model for each epoch, 
node, and layer, and
locally
stores such information in a three-dimensional tensor.
Whenever it is necessary to review an inference-time decision, the server combines distance information with the layer-wise relevance values it can compute as per \Sec{lwr}, 
and uses them to compute  
a new metric, that we named {\em relevance-aware distance} (\path{RAdist}). 

More specifically, for each sample processed through the trained model during the inference phase, 
the \path{RAdist} metric is obtained  by multiplying the aforementioned three-dimensional tensor by a 
column  vector with  as many elements as the number of layers, containing the layer-wise relevance 
corresponding to that sample. 

The result is  a matrix, containing the \path{RAdist} values for each epoch  and node.  
If the \path{RAdist} metric for a specific node, averaged over the epochs,
is significantly 
(e.g., by a factor~$\alpha$) larger than the average computed over the nodes and the epochs, 
that node is identified as misbehaving, 
and additional action can be taken -- e.g., re-training the model excluding the misbehaving node.

Thanks to the way it is defined 
and to the fact that it leverages information from {\em both the training and inference phases} of FL,
the \path{RAdist} metric is able to achieve the two main objectives of NL-FL, 
namely (i) quantifying the influence of {\em individual} learners (and their datasets) 
on the training process (unlike layer-wise relevance~\cite{bach2015pixel,Montavon2019}), 
and (ii) focusing on {\em specific} decisions, as opposed to considering all parameters~\cite{fung2020limitations}.

We stress that NL-FL differs from XML techniques %
in that it does not 
explain the decision in terms of the present, inference-phase input (step~5 of \Fig{fresco}), 
but rather in terms of past, training-phase contributions from learning nodes. 
At the same time,
an important feature of NL-FL is that it holds learning nodes liable for {\em specific} decisions concerning 
{\em individual} input samples, e.g., the classification of a given image. This is in contrast with 
similarity-based techniques to identify misbehaving learning nodes~\cite{fung2020limitations}, 
which account for the overall similarity between parameter updates. 
Thanks to this feature, NL-FL can identify misbehaving nodes also in scenarios where they only mis-label 
a small fraction of their local data.

Finally, it is worth highlighting that 
computing the \path{RAdist} metric -- and, in general, achieving the NL-FL vision -- 
entails  {\em a modest cost} in terms of additional data storage and computation.
 Indeed, we first need to compute the relevance values during the inference phase.  
As per~\cite{Montavon2019}, 
 the cost of computing the relevance values for a sample is similar to that of performing 
 a round of back-propagation during training -- actually smaller, as there is no optimization to perform.
Then, as per \Fig{nlfl}, the learning server needs to store distance and relevance information.

The former only takes up a small extra space at the learning server, 
while the latter can be quite significant, comparable to the size of the items being classified. 
For this reason, as discussed in \Sec{discussion}, it is often preferable to compute relevance values on 
an as-needed basis.

Notice that all such extra requirements brought by NL-FL affect the learning server: 
{\em no extra load} is placed on the learning nodes, which are often resource-constrained devices. 
Also, no additional operations are necessary during the {\em training} phase, which is usually 
the most resource-intensive one, and no additional data is transmitted over the network. 
In particular, NL-FL does not require learning nodes to share any additional information, 
thus the accountability brought by NL-FL does not come at 
the cost of jeopardizing FL's privacy properties.

\section{Experiment design and results}
\label{sec:results}

We now describe the behavior of the learning nodes in our experiments, 
the data at their disposal, and the DNN they run, and we present our results.

\subsection{Reference scenario and benchmarks}

We consider a typical medium-scale edge scenario~\cite{in-edge-ai}, with 10~learning nodes connected with, and coordinated by, an edge-based learning server. Each learning node has a local dataset of 4,000~images coming from the EMNIST dataset, representing handwritten digits or letters. Using the EMNIST dataset allows us to obtain meaningful, easy-to-generalize results, without the need to implement overly complex DNNs.
To evaluate \path{RAdist} in a more challenging scenario, we make local datasets non-i.i.d., by assigning to each learning node a randomly-chosen symbol (digit or letter) that is ten times more frequent than the others.
The learning nodes perform a classification task, i.e., associating each image with the character it represents. 
To this end, as in the original EMNIST paper, nodes run a DNN with four layers: 
two convolutional ones followed by two fully-connected ones. %

Nine out of the ten learning nodes are correct, i.e., all their local data are truthfully labeled. 
A tenth node misbehaves; specifically, it labels all occurrences of character~$x$ as digit~$9$. 
Such a pattern can be observed in two very different and equally relevant real-world situations:
(i) a sophisticate attack, performed by a malicious node interested in swaying the behavior of the classifier for only some of the classes, or
(ii) a honest but malfunctioning node, subject to a failure of its equipment.
Intuitively, we expect a misbehaving node mislabeling only some of its local data to be harder to detect than a node 
sending bogus, e.g., random updates. Indeed, in our case, the parameters of the misbehaving node will be very close to those of the correct nodes, hence, naive distance metrics such as those used in~\cite{fung2020limitations} may not be sufficient.

All nodes participate in a FL task, sending parameter updates to the centralized learning server after each epoch; 
the training process lasts 50~epochs in total. The server uses a simple averaging strategy to determine the global 
parameters. Further, all nodes correctly follow the FL protocol, i.e., send their (true) local parameters and dutifully replace them with the global ones upon receiving them. 
We further set~$\alpha$ to~2, leaving further study on the effect of such a parameter for future work.

\begin{figure}
\centering
\includegraphics[width=.75\columnwidth]{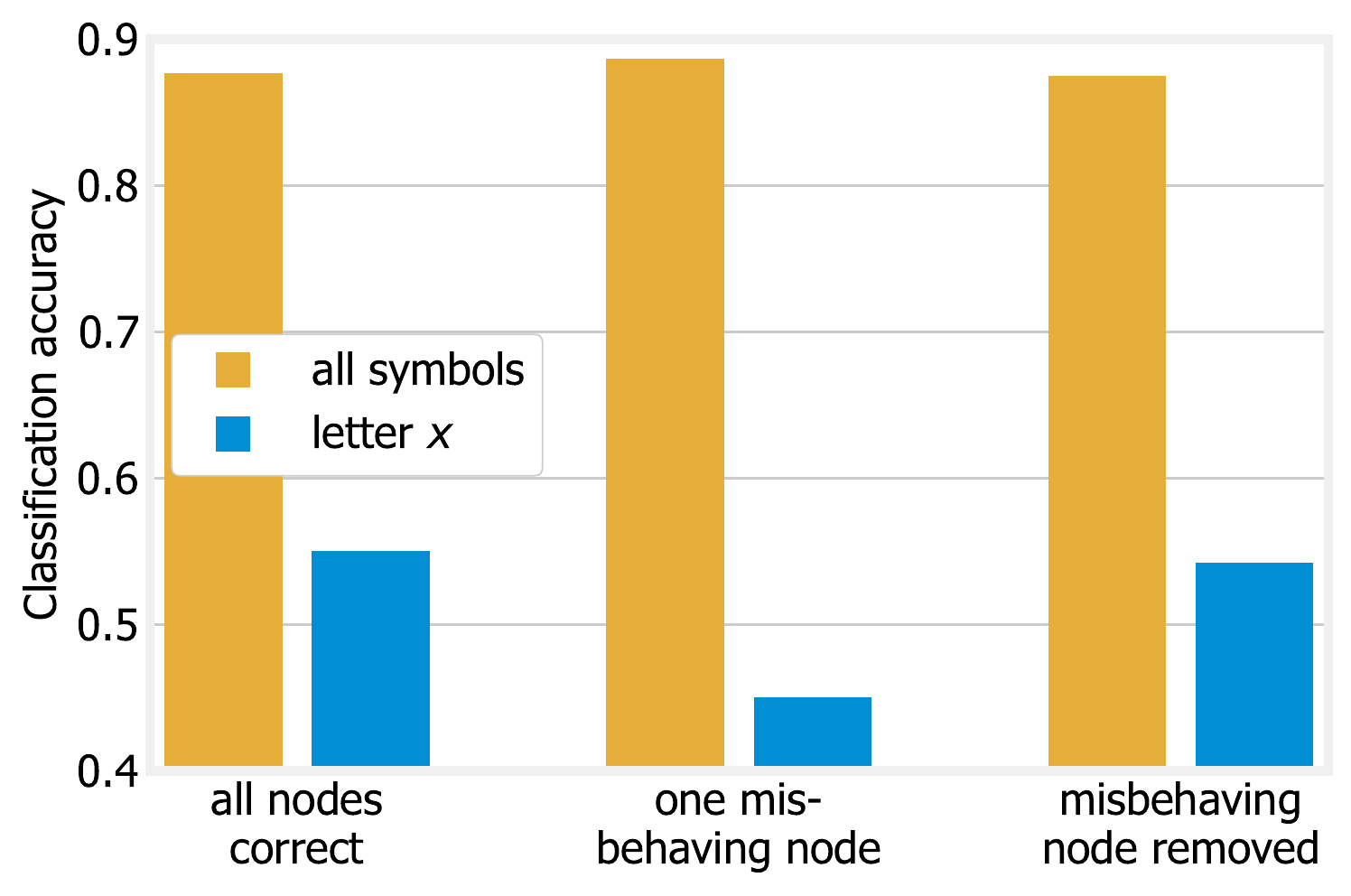}
\caption{
Accuracy for all symbols (yellow bars) and for the symbol $x$ (blue bars) when all nodes are correct (left), one node misbehaves (center), and the misbehaving node is removed (right).
\label{fig:accuracies}
} %
\vspace{-5mm}

\end{figure}

\subsection{Experimental results}

A first question we seek to answer is to which extent a single misbehaving node mislabeling a single character can affect 
the overall learning process. To this end, in \Fig{accuracies} we plot the classification accuracy for the whole dataset (yellow bars) and for the symbol~$x$ (blue bars) in three scenarios, namely: (i) when all nodes behave correctly; (ii) when one node misbehaves as detailed above, and (iii) when the misbehaving node is removed, and only the remaining nine take part in the learning.

The average accuracy is not significantly affected by the presence of the misbehaving node; after all, only a small number of images is misclassified. 
Focusing instead on the occurrences of letter~$x$ (blue bars) and comparing the first and second groups of bars, 
it is possible to see a visible drop in the accuracy, which is even more serious because letter~$x$ is not very well 
classified in the first place. Notice how such a significant effect is obtained by a single misbehaving node, 
in spite of the nine correct ones. Finally, the rightmost group of bars shows that identifying and removing 
the misbehaving node is sufficient to essentially restore the original accuracy.

\begin{figure}
\centering
\includegraphics[width=.8\columnwidth]{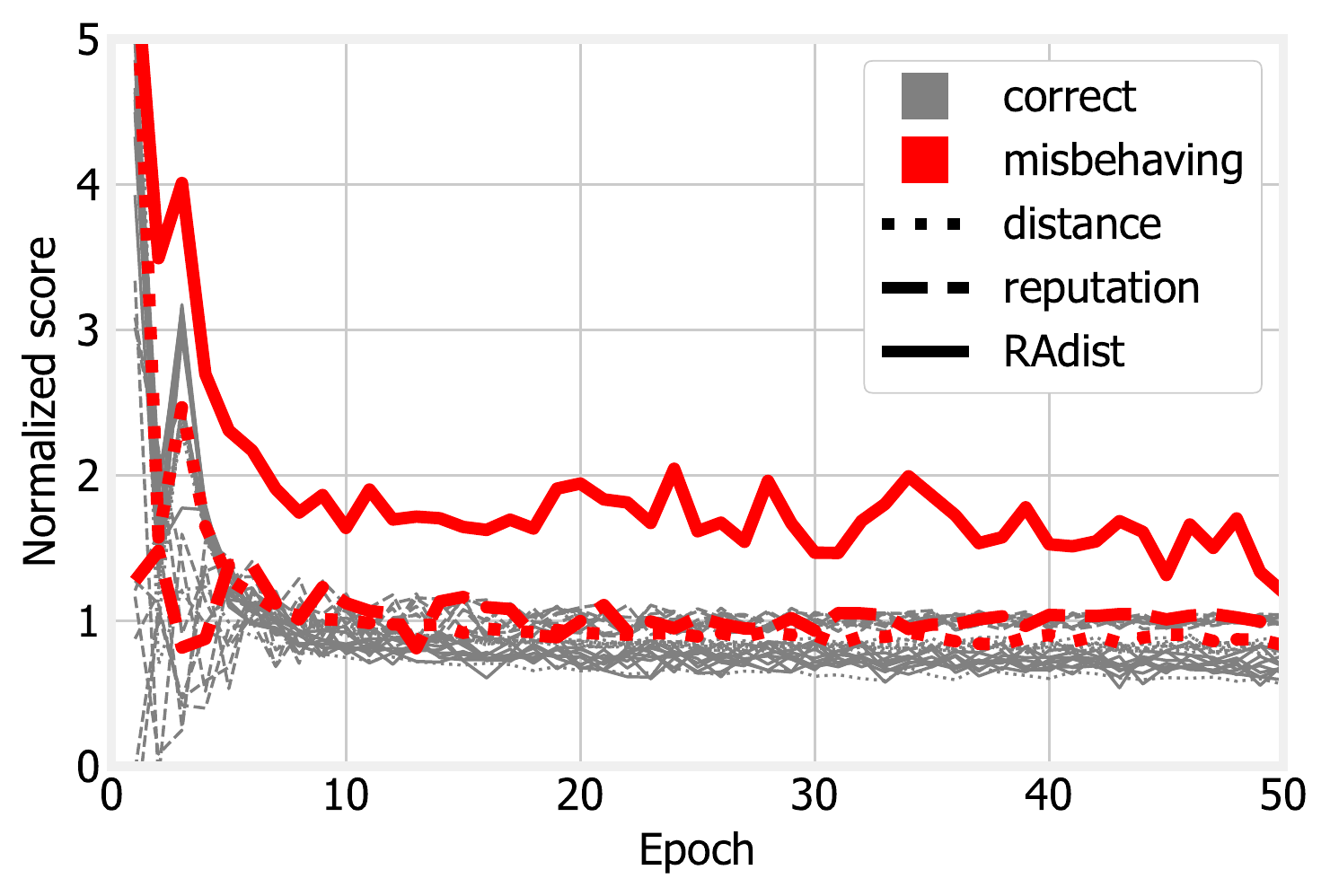}
\caption{
Scores for correct (gray lines) and misbehaving (red lines) nodes, under three scoring metrics: cosine distance~\cite{fung2020limitations} (dotted lines), reputation~\cite{kang2020reliable} (dashed lines), \protect\path{RAdist} (solid lines).
\label{fig:dists}
} %
\vspace{-5mm}

\end{figure}

We now seek to understand the usefulness of the \path{RAdist} metric in identifying such a node. 
To this end, \Fig{dists} shows the normalized {\em score} for correct (gray) and misbehaving (red) nodes, 
where ``score'' is defined as:
    (i) cosine distance~\cite{fung2020limitations} (dotted lines);
    (ii) a reputation system similar to~\cite{kang2020reliable} (dashed lines), where the reputation of a node reflects its ability to properly classify its local dataset;
    (iii) \path{RAdist} (solid lines).
The difference is very clear: dotted and dashed lines tend to lie close to each other, as both cosine distance and reputation tend to be swayed by the fact that even the misbehaving node behaves correctly {\em most of the times}, and that even correct nodes may occasionally provide wrong results.
On the other hand, when moving to the \path{RAdist} metric (solid lines), the misbehaving node emerges as having a significantly score distance than the others, for all epochs. In other words, relevance values act like a magnifying glass over the differences between parameters coming from correct and misbehaving nodes, allowing such differences to clearly emerge and drive decisions about client selection in FL.
This, in turn, allows us to neutralize the effect of the misbehaving nodes on the resulting accuracy, as shown in \Fig{accuracies}: specifically, leveraging the \path{RAdist} metric we can move from the second to the third group of bars in \Fig{accuracies}.

\begin{table}
\caption{
Overhead associated with NL-FL
\label{tab:cost}
}
\footnotesize{
\begin{tabularx}{\columnwidth}{ |X|c|c| }
  \hline
  Metric & FL & NL-FL \\
  \hline\hline
  Training time & \multicolumn{2}{c|}{27 ms/sample} \\
  \hline
  Inference time & 130 $\mu$s/sample & 153 $\mu$s/sample \\
  \hline
  Model size & \multicolumn{2}{c|}{3.41 MByte} \\
  \hline
 Data transfer rate & \multicolumn{2}{c|}{52.5 kbit/s} \\
  \hline
  Similarity info. size & -- & 7.2 kByte/epoch/node \\
  \hline
  Relevance info. size & -- & 4 kByte/sample \\
  \hline
\end{tabularx}
} %
\end{table}

Last, \Tab{cost} summarizes the overhead of NL-FL in terms of computational time, additional storage, and network latency. 
As discussed in \Sec{nlfl}, we can observe that NL-FL comes at a modest cost in terms of inference time, and the learning 
server has to store additional information concerning how close updates from different nodes are, and relevance values for 
each sample.
The space taken by similarity information grows linearly with the number of learning nodes, e.g.,  
a scenario including 1,000~learning nodes executing 100~epochs would result in 
720~MByte
of similarity information -- an acceptable overhead  for a server with sufficient capabilities 
to coordinate 1,000~nodes. The size of relevance information
can become significant for very large-scale datasets. As discussed later, its impact can be reduced by activating NL-FL only when needed.
Importantly, relevance and similarity information are created and stored locally at the coordinator, hence, neither contributes to the network overhead.
In \Tab{cost}, it is also important to observe the quantities that do {\em not} change across columns. 
Specifically, NL-FL changes neither the training performance (which is usually the most time-consuming part of learning), 
nor the network overhead. Furthermore, it places no additional burden (i.e., no extra computation or storage requirements) 
on the learning nodes.

\section{Discussion and open challenges}
\label{sec:discussion}

Through the \path{RAdist} metric, we have been able to establish a link between wrong 
decisions made during the {\em inference phase} and the nodes providing the wrong 
information during the {\em training phase} 
driving such decisions. In other words, we have been able to hold the misbehaving node liable  
for the wrong classification decisions it caused, fulfilling the basic task of NL-FL. 
At the same time, several important challenges remain open before the full NL-FL vision can be realized.

A first challenge concerns the {\em integration} of NL-FL within the wider task of
client selection and model weighting in FL. 
As an example, it is important to study the effect of parameter~$\alpha$ on the performance of NL-FL, balancing the usual trade-off between false positives and false negatives. Furthermore, we need to decide 
how wrong decisions shall be reported, and which wrong decisions trigger a liability investigation. 
Once one or more learning nodes are associated with the decisions being investigated, it is important to decide what to do with them.
As discussed earlier, removing those nodes from the learning process and repeating the training is the default action; however, if
the misbehavior is due to malfunctioning, the affected nodes can be fixed. At the same time,
some forms of misbehavior may be ground for civil or criminal liability, and need to be reported accordingly.

Another relevant issue is whether, and how, parameters coming from nodes that are unintentionally misbehaving can be 
{\em recovered} and still used for the training. Consider the scenario of \Sec{results}: we are throwing away all 
information from the misbehaving node, which still classifies images correctly in the vast majority of cases. 
It is possible that, using the \path{RAdist} metric as a guidance, parameter updates from misbehaving nodes could be amended 
and used in the learning process.
Similarly, it would be highly desirable to remove the influence from misbehaving nodes on the global model {\em without} repeating the whole training.

Finally, given the modest but nonzero overhead,  
there is the issue of deciding {\em when} 
to activate NL-FL. Specifically, instead of storing relevance values for all samples 
being classified, one can:
(i) store such information for a limited time, %
and/or (ii) require the misclassified sample when the anomaly is reported.
In the latter case, relevance values can be obtained as easily and swiftly as needed.

\section{Conclusion}
\label{sec:conclusion}

This paper introduced the concept of node liability in federated learning (NL-FL), which 
allows classification decisions to be associated with the information used at training time and their sources. 
We evaluated the NL-FL performance and overhead in an edge-based scenario, 
finding that NL-FL is much more effective than state-of-the-art solutions in 
identifying misbehaving learning nodes, 
at the cost of a modest increase in computational and storage requirements for the learning server.

\bibliographystyle{IEEEtran}
\bibliography{refs}%

\vspace{-12mm}

\begin{IEEEbiographynophoto} %
{Francesco Malandrino} (M'09, SM'19) 
is a researcher at the National Research Council of Italy (CNR-IEIIT). 
His research interests include the architecture and management of wireless, cellular, and 
vehicular networks.
\end{IEEEbiographynophoto}

\vspace{-12mm}

\begin{IEEEbiographynophoto} %
{Carla~Fabiana Chiasserini} (M'98, SM'09, F'18) 
is Full Professor  with Politecnico di Torino and a Research Associate with CNR-IEIIT. 
Her research interests include architectures, protocols, and performance analysis of wireless networks.
\end{IEEEbiographynophoto}

\end{document}